%% file: ms.tex
\documentclass[useAMS,usenatbib]{mn2e}
\usepackage{graphicx, color, setspace, ulem, natbib}
\title[Evolution of Primordial DM Haloes through Mergers]{Dynamical Evolution of Primordial Dark Matter Haloes through Mergers}
\author[Ogiya, Nagai and Ishiyama]{Go Ogiya$^{1,2}$\thanks{E-mail:ogiya@mpe.mpg.de}, Daisuke Nagai$^{3}$ and Tomoaki Ishiyama$^{4}$\\
$^{1}$Max-Planck-Institut f\"ur extraterrestrische Physik, Postfach 1312, Gie\ss enbachstra\ss e, D-85741 Garching, Germany \\
$^{2}$Universit\"ats-Sternwarte M\"unchen, Scheinerstra\ss e 1, D-81679 M\"unchen, Germany\\
$^{3}$Department of Physics, Yale University, New Haven, CT 06520, USA\\
$^{4}$Institute of Management and Information Technologies, Chiba University, 1-33 Yayoi-cho, Inage-ku, Chiba, 263-8522, Japan
}
\begin{document}

\date{Accepted 2016 June 24. Received 2016 June 23; in original form 2016 April 11}

\pagerange{\pageref{firstpage}--\pageref{lastpage}} \pubyear{2016}

\maketitle

\label{firstpage}

\begin{abstract}
Primordial dark matter (DM) haloes are the smallest gravitationally bound DM structures from which the first stars, black holes, and galaxies form and grow in the early universe. However, their structures are sensitive to the free streaming scale of DM, which in turn depends on the nature of DM particles. In this work, we test the hypothesis that the slope of the central cusps in primordial DM haloes near the free streaming scale depends on the nature of merging process.  By combining and analysing data from a cosmological simulation with the cutoff in the small-scale matter power spectrum as well as a suite of controlled, high-resolution simulations of binary mergers, we find that (1) the primordial DM haloes form preferentially through major mergers in radial orbits; (2) their central DM density profile is more susceptible to a merging process compared to that of galaxy and cluster-size DM haloes; (3) consecutive major mergers drive the central density slope to approach the {\it universal} form characterized by the NFW profile, which is shown to be robust to the impacts of mergers and serves an attractor solution for the density structure of DM haloes. Our work highlights the importance of dynamical processes on the structure formation during the Dark Ages.
\end{abstract}


\section{Introduction}
\label{sec:int}

In the hierarchical structure formation model, primordial dark matter (DM) haloes are the smallest gravitationally bound DM structure from which the first stars, black holes, and galaxies form and grow in the early universe. These small, ancient structures are potential targets for upcoming observations with {\it James Webb Space Telescope} (JWST), {\it Giant Magellan Telescope} (GMT), {\it Thirty Meter Telescope} (TMT), and {\it European Extremely Large Telescope} (E-ELT) \citep[for a recent review, see e.g.][]{2011ARA&A..49..373B} and hold promise to shed light on the structure formation in the so-called Dark Ages. 

However, the structure of primordial DM haloes are sensitive to the free streaming scale of DM, which in turn depends on the mass of DM particles. 
Below the free streaming scale, formation of smaller scale structure is suppressed and DM haloes form through direct gravitational collapse instead of mergers of smaller objects. As such, the structure of DM haloes near the free streaming scale are likely very different from the DM haloes on larger (e.g., galactic and galaxy cluster) scales. For example, DM haloes near the free streaming scale have earlier formation epochs and form denser cores than their more massive counterparts. Primordial DM haloes are also expected to be abundant and ubiquitous \citep[e.g.][]{2005Natur.433..389D}, making them potential sources of the DM annihilation signals (\citealp[e.g.,][]{2003PhRvD..68j3003B,2007JCAP...04..016B,2008PhRvD..77h3519B}, but see also \citealt{2008Natur.456...73S}). 

One of the leading candidates of cold dark matter (CDM), neutralino, is the lightest supersymmetric particle with mass of order $m_{\rm \chi} \sim 100 {\rm GeV}$. This finite streaming scale sets the lower mass limit of DM haloes to be around the Earth mass ($\sim M_{\rm \oplus} \sim 10^{-6} M_{\rm \odot}$) and introduces the cutoff in the matter power spectrum below the streaming scale (\citealp[e.g.,][]{2004MNRAS.353L..23G}, but see also \citealt{1999PhLA..260..262Z,2006PhRvL..97c1301P,2009NJPh...11j5027B}). At present, particle accelerator experiments have placed a lower limit on the neutralino mass, $m_{\rm \chi} > 37 {\rm GeV}$ \citep{2002PhRvD..66a0001H}, while its upper limit is $m_{\rm \chi} \la 500 {\rm GeV}$ from the cosmic matter density measured by the {\it Wilkinson Microwave Anisotropy Probe} (WMAP) \citep{2003PhLB..565..176E}.

Numerous cosmological $N$-body simulations without the cutoff in the small-scale matter power spectrum have been carried out to investigate the density structure of DM haloes on the scales in which the effects of the free streaming damping is negligible. 
\citet[][NFW]{1997ApJ...490..493N} found that the density structure of CDM haloes is almost universal over a wide range of halo mass, from galactic to cluster scales, and is well described by the NFW profile, 
\begin{equation}
\rho(r) = \frac{\rho_{\rm s}}{(r / r_{\rm s}) [1 + (r / r_{\rm s})]^{2}}, \label{eq:nfw} 
\end{equation}
where $r$ represents the distance from the centre of the halo and $\rho_{\rm s}$ and $r_{\rm s}$ are the scale density and length, respectively. The concentration parameter of the DM haloes is defined as the ratio between the virial radius, $r_{\rm vir}$, and the scale length, $r_{\rm s}$, of haloes, $c \equiv r_{\rm vir}/r_{\rm s}$.
Subsequent studies confirmed the results of NFW and also showed deviations from the universality \citep[e.g.][]{1997ApJ...477L...9F, 1999MNRAS.310.1147M, 2000ApJ...529L..69J}. 
Recent work with higher resolutions indicate that the Einasto profile \citep{1965TrAlm...5...87E} provides better fits to simulation results \citep{2008MNRAS.391.1685S, 2009MNRAS.398L..21S, 2010MNRAS.402...21N, 2013ApJ...767..146I, 2014MNRAS.441.3359D}, especially in the central region ($r < 0.01 r_{\rm vir}$). 

The structure of DM haloes in models with the cutoff in the small-scale matter power spectrum has been more controversial. 
Analytic work predicted that the truncated power spectrum forms density cores in the centre of individual DM haloes \citep{2004ApJ...604...18W, 2012MNRAS.424L...6V}. 
However, cosmological $N$-body simulations with the cutoff in the small-scale matter power spectrum obtained DM haloes whose density structure is well fitted by the NFW profile \citep[e.g.][]{2007ApJ...665....1B, 2009MNRAS.396..709W}. 
Recent $N$-body simulations suggest that the dynamics and evolution of DM haloes may play an important role in determining the density structure of the primordial DM haloes near the free streaming scale and found that the central DM density cusps are steeper than the NFW profile, i.e. $\rho \propto r^{-\alpha}$ 
\citep[$\alpha > 1$;][]{2010ApJ...723L.195I, 2013JCAP...04..009A}. 
For example, \citet[][hereafter I14]{2014ApJ...788...27I} and \cite{2016arXiv160403131A} showed that the central density profiles become shallower through mergers in DM haloes forming near the free streaming scale. These results are in contrast to the results of dissipationless mergers of two isolated DM haloes, which suggested strong preservation of their initial density structure \citep{2004MNRAS.349.1117B, 2006ApJ...641..647K, 2008MNRAS.386.1543Z}.  

Another potential candidate of DM includes warm dark matter (WDM) which has larger free streaming scales than the CDM. WDM is expected to have particle masses of order $1-10 {\rm keV}$, and the lower mass limit of WDM haloes corresponds to that of dwarf galaxies with mass of about $10^{8-9} M_{\rm \odot}$ \citep{2001ApJ...556...93B}. A recently detected unidentified emission line of $E \approx 3.55 {\rm keV}$ from M31 and galaxy clusters, for example, may be decay signals of a kind of WDM candidates, sterile neutrino \citep{2014PhRvL.113y1301B, 2014ApJ...789...13B}. Although their free streaming scales are quite different from that of CDM, there are some similarities on the impacts of finite free streaming scale on the structure of DM haloes \citep[e.g.][]{2012MNRAS.424.1105M, 2013MNRAS.434.3337A, 2013MNRAS.430.2346S, 2014MNRAS.439..300L, 2015MNRAS.450.2172P}, suggesting that the results based on CDM with finite streaming scale may provide useful insights into the structure formation processes in the WDM models.

In this work, we investigate the dynamical impacts of mergers on the density structure of primordial DM haloes formed at the early stage of the structure formation. Focusing on the DM haloes near the free streaming scale, we demonstrate the importance of a major merger in determining the structure of the primordial DM haloes. In particular, we highlight that the DM haloes near the free streaming scale at redshift $z=32$ have considerably smaller ($c \sim 2$; see Figure 8 of I14) concentration parameters than that ($c \sim 10$) of the galactic-size haloes \citep[e.g.][]{2001MNRAS.321..559B, 2012MNRAS.423.3018P}. 
The evolution of the concentration parameter, especially near the free streaming scale, remains an open question and has been a focus of recent studies \citep[e.g.][]{2015MNRAS.451.3117S,2016MNRAS.tmp..813L}. The important update of this paper therefore lies in exploration of the dynamical impact of mergers on the structure of the DM haloes with a smaller concentration parameter that have not been explored in the literature.

This paper is organized as follows. 
Section 2 studies the properties of mergers between primordial DM haloes with data from the cosmological simulations performed by I14. 
We find that the slope of the density cusps in the centre of primordial DM haloes formed through mergers are shallower than those of haloes which have not experienced mergers. 
Section \ref{sec:cont_sims} investigates the dynamical impacts of major mergers on the structure of primordial DM haloes using a suite of controlled $N$-body binary merger simulations. 
We summarize and discuss the results in Section \ref{sec:sum}. 

\section{Mergers in the cosmological simulations}
\label{sec:cosmo}

In this Section, we study the properties of mergers between primordial DM haloes extracted from a set of cosmological simulations. 
The first set, the model A, is a smaller volume version of one named A\_N4096L400 in I14, where the sharp cutoff was imposed in the power spectrum by the free streaming damping of DM particles with a mass of 100 GeV and the corresponding mass scale of the smallest DM haloes is $\sim 10^{-6} M_{\rm \odot}$ \citep{2004MNRAS.353L..23G}. 
The second set, the model B, is the cosmological simulation without the cutoff , and it is labelled B\_N2048L200 in I14. 
The simulations adopt the WMAP 7 cosmological parameter set, $\Omega_0 = 0.27$, $\lambda_0 = 0.73$, $h = 0.7$ and $\sigma_8 = 0.8$ \citep{2011ApJS..192...18K}, where the symbols have their usual meaning. 
We simulate the motion of $2048^3$ particles from $z=400$ to $z=32$ in a comoving box with a side length of 200 pc using the massively parallelized TreePM code, GreeM \citep{2009PASJ...61.1319I, 2012arXiv1211.4406I}. 
The Plummer softening length and mass resolution are $2 \times 10^{-4} {\rm pc}$ and $3.4 \times 10^{-11} M_{\odot}$, respectively. 
These values are the same as those used in I14 ($4096^3$ particles, 400pc).
We refer readers to I14 for further details on the numerical simulations.

DM haloes are identified by using the ROCKSTAR halo/subhalo finder \citep{2013ApJ...762..109B} from 11 snapshots output at $z=52, 50, 48, 46, 44, 42, 40, 38, 36, 34$ and 32, and the gravitationally consistent merger tree has been constructed \citep{2013ApJ...763...18B}.  The orbital parameters of mergers are derived using the procedure described in \cite{2011MNRAS.412...49W}. 

\subsection{Parameters of mergers between primordial DM haloes}
\label{sec:prop_mrgs}
\input{fig1.tex}
Figure \ref{fig:mrg_params} shows the distribution of the parameters of mergers in which the virial mass of primal DM haloes, $M_{\rm vir, pri} \geq 7.14 \times 10^{-7} M_{\rm \odot}$, where we use virial mass and radius defined by \cite{1998ApJ...495...80B} throughout this paper. 
The resolvable mass scale of DM haloes determined by the mass resolution of the simulation corresponds to the mass of the smallest haloes in the model B (the case without the cutoff). 
We therefore analyse the model B using two sets of minimum subhalo masses with $M_{\rm sub, min}=7.14 \times 10^{-7}$ and $7.14 \times 10^{-8} M_{\rm \odot}$. The latter would be more suitable for the model B since there is no physical reason to introduce the lower mass limit of DM haloes, while we adopt $M_{\rm sub, min}=7.14 \times 10^{-7} M_{\rm \odot}$ for the model A. 

The top panel shows the distribution of the orbital circularity, $\eta \equiv L / L_{\rm c}$, defined as the ratio between the angular momentum, $L$, and that of the circular orbit, $L_{\rm c}$.  Although there is a relatively wide range of $\eta$, we find that the primordial DM haloes merge with more radial orbits (i.e., small circularities $\eta$) in the model A (red boxes, with cutoff) than in the model B (black boxes, without cutoff).
For example, $\eta = 0.3$ (equivalently the orbital eccentricity of $e = \sqrt{1-\eta^2} = 0.95$) is most common in the model A, while $\eta \sim 0.5$ is more common in the model B. Note that the difference between the two models becomes more pronounced when we adopt $M_{\rm sub, min} = 7.14 \times 10^{-8} M_{\rm \odot}$ for the model B (black dashed).

The middle panel depicts the distribution of $R \equiv r_{\rm apo} / r_{\rm vir}$, the ratio of the apocentre, $r_{\rm apo}$, and the virial radius of the primal haloes, $r_{\rm vir}$. The majority of mergers has apocentre of  $R \sim 1$ in the both models.

The bottom panel shows the distribution of the ratio between virial masses of primal- and subhaloes, $M \equiv M_{\rm vir, pri} / M_{\rm vir, sub}$, showing that major mergers ($M \sim 1$) are more dominant in the model A compared to the model B.
Major mergers are also dominant in the model B for $M_{\rm sub, min} = 7.14 \times 10^{-7} M_{\rm \odot}$ (black solid), but minor mergers ($M \sim 10$) become more dominant for the case of $M_{\rm sub, min} = 7.14 \times 10^{-8} M_{\rm \odot}$ (black dashed). 

Compared to the orbital parameters of galaxy or cluster-size DM haloes which typically have the orbital circularity of $\eta = 0.5$ \citep[e.g.][]{2006A&A...445..403K, 2008ApJ...675.1095J, 2011MNRAS.412...49W}, the primordial DM haloes merge with more radial orbits in the model with a finite streaming scale, as shown in the top panel of Figure \ref{fig:mrg_params}. Interestingly, mergers between primordial DM haloes with $\eta \sim 0.5$ are typical in the model B, similar to the results of galaxy and cluster-size DM haloes formed during the late phase of the structure formation.

\cite{2011MNRAS.412...49W} showed that merging orbits become more radial and plunge deeper into primal haloes at higher redshift and for primal haloes with higher mass, and proposed a fitting formula of the distribution of orbital parameters of mergers as a function of redshift, $z$, and mass of primal haloes, $M_{\rm pri}$. 
According to this formula, 40\% of mergers of DM haloes have apocentres of $0.3 \leq R \leq 3$ 
for the case of $M_{\rm pri}=10^{12}M_{\rm \odot}$ at $z = 0$ \citep[see also][]{2012MNRAS.424..361M}, suggesting that $R = 1$ is typical for galaxy-sized DM haloes. 

Since smaller haloes are more abundant on the scales of galaxies or galaxy clusters, minor mergers occur more frequently than major mergers. 
However, a larger fraction of major mergers is expected for DM haloes near the free streaming scale because of the absence of substructures. 
The result shown in the bottom panel of Figure \ref{fig:mrg_params} is therefore consistent with this expectation. The analysis of the model B indicates that the fraction of minor mergers is dominant even in the early phase of the structure formation for the case without the cutoff.

\subsection{Mergers and density structures of primordial DM haloes}
\label{sec:mrgs_dens}

\input{fig2.tex}
We test the hypothesis that the slope of the central cusps in primordial DM haloes near the free streaming scale depends on the nature of merging process.
Figure \ref{fig:slope_cosmo} demonstrates the logarithmic slope of the stacked density profile of primordial DM haloes with virial masses $1 \times 10^{-6} \leq M_{\rm vir}/M_{\rm\odot} \leq 4 \times 10^{-6}$ at $z=32$. In order to make a sample set of DM haloes which are bound and stable, we impose the condition, $0.05 \leq K/|W| \leq 0.95$, where $K$ and $W$ are kinetic and potential energies of each halo. The average virial mass of the sample haloes is $\sim 1.5 \times 10^{-6} M_{\rm \odot}$. Using the merger tree, we identify mergers with $M_{\rm vir, pri} \geq M_{\rm vir, sub} \geq 7.14 \times 10^{-7} M_{\rm \odot}$, where $M_{\rm vir, pri}$ and $M_{\rm vir, sub}$ are the virial masses of the primal- and subhaloes, respectively.
Red and blue lines are the results for primordial DM haloes which have not and have experienced mergers by $z=32$, respectively.
Compared to the prediction by I14 (indicated by the black line), we find that DM haloes which have experienced no merger (at least one merger) have steeper (shallower) cusps, suggesting that mergers play an important role in making the inner density profile of the primordial DM haloes shallower.

\section{Controlled Simulations}
\label{sec:cont_sims}
We study the dynamical impacts of mergers on primordial DM haloes by performing and using a suite of high-resolution, controlled $N$-body binary merger simulations.

\subsection{Set up}
\label{sec:setup}
Initial positions and velocities of particles in $N$-body systems are determined by the method proposed by \cite{2006ApJ...641..647K}. 
This generates $N$-body systems in the equilibrium states. 
The density structure of DM haloes within the virial radius, $r_{\rm vir}$, initially follows 
\begin{equation}
\rho(r) = \rho_{\rm in}(r) = \frac{\rho_{\rm s}}{(r / r_{\rm s})^{\alpha} [1 + (r / r_{\rm s})]^{\beta - \alpha}} \ (r \leq r_{\rm vir}), \label{eq:inside} 
\end{equation}
where $\alpha$ and $\beta$ are the logarithmic slopes of the central density cusp and outskirt, respectively. 
We assume $\beta = 3$ and $c = 2$ throughout this Section. 
The models with $\alpha = 1$ correspond to the NFW profile.

However, the mass profile of the models with $\beta \leq 3$ diverges as $r \rightarrow \infty$. 
In order to ensure the finite mass of individual systems, we adopt an exponentially decaying density structure beyond $r_{\rm vir}$ as 
\begin{equation}
\rho(r) = \rho_{\rm in}(r_{\rm vir}) \biggl ( \frac{r}{r_{\rm vir}} \biggr )^{\kappa} \exp{\biggl (- \frac{r - r_{\rm vir}}{r_{\rm decay}} \biggr )} \ (r > r_{\rm vir}). \label{eq:outside}
\end{equation}
The decaying radius, $r_{\rm decay}$, controls the truncation of the particle distribution and the sharpness of the transition of the density structure. 
Requiring the logarithmic slope to be continuous at $r = r_{\rm vir}$, a slope in Equation (\ref{eq:outside}), $\kappa$, is given by 
\begin{equation}
\kappa = -\frac{\alpha + \beta c}{1+c} + \frac{r_{\rm vir}}{r_{\rm decay}},
\end{equation}
where $r_{\rm decay}$ is determined by imposing $M(< r_{\rm vir}) \equiv M_{\rm vir} = F M_{\rm tot}$, $M_{\rm tot}$ is the total mass of the system, and we adopt $F = 0.8$ for all runs. 

\input{fig3.tex}
In order to study the dynamical impacts of major mergers, we prepare the systems in the equilibrium states initially. We also assume that the velocity dispersion of the systems is initially isotropic and that the phase-space distribution function depends only on energy. In order to assess the effects of artificial two-body relaxation and/or insufficient force resolutions, we follow the prescription outlined in \citet[][P03]{2003MNRAS.338...14P} \citep[see also][]{2001ApJ...557..533F}. 
According to P03, simulation results are reasonable when  
\begin{equation}
T_{\rm relax}(r) \equiv \frac{N(<r)}{8 \ln{N(<r)}} T_{\rm d}(r) > 0.6 T_{\rm int} \label{eq:trelax}
\end{equation}
is satisfied, where 
$T_{\rm relax}$ and $T_{\rm int}$ are the timescales of two-body relaxation and integration time, respectively. 
The free-fall time measured at $r$, $T_{\rm d}(r)$, is defined as 
\begin{equation}
T_{\rm d}(r) \equiv \sqrt{\frac{r^3}{G M(<r)}}, \label{eq:td}
\end{equation}
where $G$ is the Newton's gravitational constant. 

P03 also showed that the resolvable range of the system satisfies the condition, 
\begin{equation}
a_{\rm char} \geq a_{\rm exp}(r) = \frac{GM(<r)}{r^2}, \label{eq:aexp}
\end{equation}
where the characteristic resolvable acceleration, $a_{\rm char}$, is defined as 
\begin{equation}
a_{\rm char} \equiv \frac{\chi G M_{\rm vir}}{r_{\rm vir} \epsilon}, \label{eq:achar}
\end{equation}
where $\epsilon$ is the softening parameter. 
They empirically obtained $\chi = 0.5$. 
P03 proposed an expression to determine the optimal softening parameter for systems with given numbers of particles within the virial radius, $N(< r_{\rm vir})$, 
\begin{equation}
\epsilon = \frac{A r_{\rm vir}}{\sqrt{N(< r_{\rm vir})}}. \label{eq:epsilon}
\end{equation}
To ensure the features of collisionless systems, $A$ must be equal or greater than unity. 
P03 found that $A = 4$ provides sufficient force resolutions and minimized numbers of timesteps for haloes in cosmological simulations 
without the cutoff in the power spectrum, i.e. NFW haloes ($\alpha = 1.0$). 
However, haloes with steeper cusps ($\alpha > 1.0$) may require higher force resolutions ($1 \leq A < 4$) because of the stronger gravitational acceleration in their central region. 

Figure \ref{fig:res_study} shows the stability tests and required force resolutions for models with $\alpha=1.0$ (left), $\alpha=1.5$ (middle) and $\alpha=2.0$ (right). In this figure, we simulate the dynamical evolution of isolated $N$-body systems for varying force resolution parameters, $A = 1$ (red), $2$ (blue) and $4$ (magenta). 
The effects of two-body relaxation can be ignored in the range where $(5/3) T_{\rm relax}(r) > T_{\rm int}$ is satisfied, indicated by the radial range where black lines are above green lines in the panels of the first row. 
In this region, the systems should retain the initial configurations when the force resolution is sufficient. 
Panels in the first row also indicate that the effects of two-body relaxation are negligible, suggesting that our simulation results are trustable at $r \geq 0.01 r_{\rm vir}$ at $T_{\rm int} = 10 T_{\rm d}(r_{\rm vir})$. 
Panels in the second row compare the expected acceleration, $a_{\rm exp}(r)$, with the characteristic acceleration, $a_{\rm char}(A)$, and show that $A=4$ provides a sufficient force resolution for systems characterized by the NFW profile ($\alpha=1.0$) as shown by P03. 
Models with a very steep cusp ($\alpha=2.0$), on the other hand, require higher force resolutions in the centre of the system. 
The panels in the third and fourth rows demonstrate that systems retain their initial configuration and the results are reasonable in the range where both the conditions for the two-body relaxation timescale and force resolution are satisfied as expected. 
In the regions with insufficient force resolutions, artificial density cores have formed, and the velocity dispersion profiles deviate from the initial condition in a short time (see blue and magenta thin lines in the density and velocity dispersion profiles of $N$-body systems with $\alpha = 2.0$). We therefore only show the trustable radial range ($r \geq 0.01 r_{\rm vir}$) and adopt $A = 4$ for models of $\alpha=1.0$ and $A = 1$ for models of $\alpha = 1.5$ and $2.0$ in the following parts of this Section. 

\input{./fig4.tex}
Next, we simulate gas-less binary mergers between two identical $N$-body systems with high resolutions. 
Each DM halo is represented by $N$ particles. Since the two systems have the same mass, all particles have equal masses in each simulation and the total number of particles is $N_{\rm tot} = 2 N$ in merger simulations. 
Two DM haloes are initially located at the apocentre of the relative orbit separated by a distance $d = D r_{\rm vir}$, where the dimensionless parameter, $D$, controls the separation between the centres of the two systems. 
The relative velocity is $V_{\rm ini} = \eta V_{\rm c}(d) = \eta [GM(<d) / d]^{1/2}$, where $M(<d)$ is the mass of a merger progenitor enclosed in $d$. 
The orbit is confined on the XY plane by assuming the merger of two systems with extended mass distributions as one between two point masses. 
The centre of the systems at given time is defined as the point of the potential minimum. 

\input{./fig5.tex}
Simulations are performed with the tree code \citep{1986Natur.324..446B} which adopts the second-order Runge-Kutta scheme in time integration and designed for graphic processing unit (GPU) clusters. 
The division of roles between CPU cores and GPU cards follows \cite{2011arXiv1112.4539N}. 
CPU cores construct oct-tree structures of particles and GPU cards calculate gravitational acceleration through traversing the tree structures \citep{2013JPhCS.454a2014O}. 
The opening angle of the tree algorithm is set to be $\theta = 0.6$ in all runs. 

\subsection{Dependence on the internal structure}
\label{sec:dep_str}

\input{table1.tex}

Using the setup described in Section~\ref{sec:setup}, we first investigate the dependence of the dynamical impacts of major mergers on the internal structure of progenitors by performing and analyzing a set of simulations in which we vary the central density slope, $\alpha$. Table \ref{tab:run_str} presents the parameters of these runs.

Figure \ref{fig:dns_prf} shows the spherically averaged density profiles of remnant DM haloes after major mergers. 
The density in iso-runs (green and black dashed lines) is doubled. 
These lines represent the stability of the system at $ r \leq r_{\rm vir}$. 
The merger remnants (red lines) have a higher central concentration and more extended envelope than their progenitors as shown by early studies \citep[e.g.][]{1978MNRAS.184..185W, 2004MNRAS.349.1117B}. 

The upper panels of Figure \ref{fig:dns_enh} show the ratio of the density profile of the merger remnants to the initial density profiles of their progenitors. 
The peaks of the density enhancement occur around $0.4 r_{\rm vir}$, while keeping the density around $r_{\rm vir}$ unchanged.
In the major merger between DM haloes with $\alpha=1.0$ (upper left panel), the mass density increases by more than a factor of 2 at $r < 0.6 r_{\rm vir}$. In the cases of DM haloes with steeper central cusps with $\alpha \ge 1.5$, the central density at $r < 0.1 r_{\rm vir}$ increases by $\sim 50 \%$ 
(see the upper middle and right panels). 
The logarithmic slope of the radial density profile, ${\rm d} \ln{\rho(r)} / {\rm d}\ln{r}$, becomes larger in the centre of haloes (i.e. shallowing cusps) through a major merger because the amount of the density enhancement around the centre is smaller than that in the outskirts (lower panels). 
Since DM haloes with steeper central cusps ($\alpha \ge 1.5$) exhibit the stronger radial dependence in the efficiency of the density enhancement, there are larger changes in the logarithmic slope compared to those with shallower central cusp (e.g., $\alpha=1.0$). 

\input{./fig6.tex}
To gain insights into the physical picture, Figure \ref{fig:pseudo_vir_ratio} shows the radial profiles of the pseudo virial ratio, $\sigma^2(r) / |\Phi(r)|$, where $\sigma(r)$ and $\Phi(r)$ are the radial profiles of three-dimensional velocity dispersion and gravitational potential, respectively. 
Figure \ref{fig:pseudo_vir_ratio} depicts that the central part of DM haloes with steeper cusps are dynamically hotter, i.e. higher pseudo virial ratio than that of DM haloes with shallower cusps. 
Major mergers lead to significant changes in the gravitational potential and particles exchange their energy through the process of violent relaxation \citep{1967MNRAS.136..101L}. 
In the mergers between DM haloes with $\alpha = 1.5$ or $2.0$, some particles around the centre of the original haloes gain sufficient energy to escape from their centre (see blue and magenta lines in the upper middle and right panels of Figure \ref{fig:dns_enh}) and suppress the growth of mass density through mergers, while these effects are less efficient in the mergers between DM haloes with $\alpha = 1.0$ because the central part of the progenitor haloes is dynamically cold. 

\input{./fig7.tex}
We find that the density enhancement profiles shown in Figure \ref{fig:dns_enh} arises from the quasi-stable wave structures. 
Figure \ref{fig:dns_field} illustrates the enhancement in the column density field projected on the XY-plane. Symmetrical structures are present in all cases. They are also rotating because of the angular momentum transported from the initial relative orbit and are quasi-stable (i.e., measurable at least for $5 T_{\rm d}(r_{\rm vir})$). 
The direction of the significant density enhancement is determined by that of the collision between the progenitors. 
In the most violent phase of the merger process, particles are strongly accelerated in the direction of the merger axis, causing orbits of a large fraction of particles to expand and thereby creating the peak of the density enhancement around $0.4 r_{\rm vir}$. 
In addition, spherically expanding shells appear in the inner side of the symmetrical density enhancements.
During the process of violent relaxation, a fraction of particles escapes from the centre and creates the waves around the centre in the radial profile of the density enhancement (see $r < 0.2 r_{\rm vir}$ in the upper right panel of Figure \ref{fig:dns_enh}).
Such structures are more pronounced in the mergers between DM haloes with steeper cusps which are dynamically hotter than others as illustrated in Figure \ref{fig:pseudo_vir_ratio}.

\subsection{Dependence on the orbital parameters}
\label{sec:dep_orb}

\input{table2.tex}
\input{./fig8.tex}

As illustrated in the top (middle) panel of Figure \ref{fig:mrg_params}, the cosmological simulation with the cutoff in the small-scale matter power spectrum predicts a relatively wide range of orbital circularity, $\eta$ (apocentre, $r_{\rm apo}$). To examine the dependences of the density enhancement and logarithmic density slope on orbital parameters, we perform controlled simulations described in Table \ref{tab:run_orb} in which we vary the orbital circularity, $\eta$, and the initial separation between the merger progenitors, $d= D r_{\rm vir}$.

Figure \ref{fig:dns_enh_eta} shows that the density enhancement (upper panel) and logarithmic density slope (lower panel) around the centre of the merger remnants do not depend sensitively on $\eta$. 
In the mergers with high $\eta$ ($\eta \geq 0.7$), the peak of the density enhancement occurs at the larger halo radii and the amplitude of the peak is less than those in lower $\eta$ cases. In these high $\eta$ cases, the larger amounts of orbital energy and angular momentum are transported to the merger remnant and makes it more spatially extended. The density enhancements of the merging orbits with $\eta \leq 0.5$, which are common in the cosmological simulation, are almost identical. 
Since the differences between the density enhancements in the centre and outskirts do not strongly depend on $\eta$, the merger remnants have similar profiles of the logarithmic density slope.

\input{./fig9.tex}
The upper panel of Figure \ref{fig:dns_enh_r} shows the density enhancement. 
In the case of low $D$ value ($D=0.5$; red line), the fluctuations in the gravitational potential caused by the major merger is smaller than those in the cases with higher $D$ value. 
This is because the amounts of transported orbital energy and angular momentum from the initial relative orbit are smaller. 
As a result, most of the particles around the centre of the haloes do not gain sufficient energy to escape from the central part, causing the density peak to occur at the smaller radii while enhancing the amplitude of the density enhancement compared to the standard case ($D=1.0$; blue line).
The merger with high $D$ ($D=2.0$; magenta line), on the other hand, transports the larger amounts of orbital energy and angular momentum to the merger remnants compared to the cases with smaller $D$. The merger remnant is therefore more spatially extended, causing the peak of the density enhancement to move toward larger halo radii while reducing the amplitude of the density enhancement compared to the standard case. 
These results suggest that the density enhancement around the centre of the merger remnants depends on the initial separation between two merger progenitors. 

The lower panel of Figure \ref{fig:dns_enh_r} demonstrates that the profile of the logarithmic slope depends on $D$ at $r/r_{\rm vir} \geq 0.03$ reflecting the difference in the density enhancement, but the slope in the central region ($r/r_{\rm vir} \leq 0.03$) is almost independent of $D$, interestingly.

\subsection{Impacts of consecutive major mergers}
\label{sec:consecutive}

\input{./table3.tex}
Based on the insights gained from these simulations, we develop a simple model to describe the formation of DM haloes with 
$5 \times 10^{-5} M_{\rm \odot}$ shown in Figure 3 of I14. \footnote{The upper panels of Figure 3 in I14 demonstrate the density profiles of primordial DM haloes with $2 \times 10^{-6} M_{\rm \odot}$ (left panel) and with $5 \times 10^{-5} M_{\rm \odot}$ (right panel), while the lower left panel illustrates that the logarithmic slope of the density profile of DM haloes with $2 \times 10^{-6} M_{\rm \odot}$ is $|d \ln{\rho} / d \ln{r}| \approx 1.5$ around the centre.} First, we suppose that the density of DM haloes increases by $50 \%$ at $r \leq 0.1 r_{\rm vir}$ and doubles at $r \geq 0.1 r_{\rm vir}$ through a major merger. 
This assumption is approximately consistent with the results of our controlled simulations for the models of $\alpha=1.5$.  
Assuming that DM haloes increase their mass only through major mergers, the mass growth by a factor of 25 requires about 4-5 major mergers. Since $r_{\rm vir} \propto M_{\rm vir}^{1/3}$, the mass growth by a factor of 25 causes the increase in the virial radius, $r_{\rm vir}$, by a factor of $\sim 3$. 
Table \ref{tab:app} represents the results of this simple model, which can explain the evolution of the density structures of the primordial DM haloes in the cosmological simulations. 

A recent work by \cite{2016arXiv160403131A} investigated the formation and evolution of DM haloes near the free streaming scale of the CDM using simulations analogous to ours and obtained consistent results: i.e., primordial DM haloes first emerge with steeper central cusps than that of the NFW profile and the cusps become shallower through major mergers. 
They also demonstrated that the density structure of DM haloes is well-described by the NFW profile on average when their virial mass reaches a few percent of the solar mass, consistent with the prediction of I14. Such DM haloes should have experienced multiple repeated mergers. But, the central density cusp of individual DM halos can deviate from the NFW profile depending on the details of mass accretion histories and merger parameters. Our simulations, for example, suggest the possibility of emerging the density profile with the shallower cusp as well (see the left panels of Figure \ref{fig:dns_enh}). 
Mixing in the phase-space by violent changes in the gravitational potential during major mergers likely plays an important role in the emergence of the NFW profile or ones with shallower cusps as indicated by analytic studies \citep[e.g.][]{2001ApJ...563..483T, 2013MNRAS.430..121P}. 

\input{table4.tex}
The above prediction and suggestion highlight the importance of consecutive major mergers on the dynamical evolution of the central density structure of primordial DM haloes. 
To study the dynamical impacts of consecutive mergers, we perform additional controlled simulations (see Table \ref{tab:run_cum} for a summary). 
In the simulations of the second merger labeled `2nd', the merger remnant of the snapshot at $t = 5 T_{\rm d}(r_{\rm vir})$ in the runs labeled `nr' merges with the copied identical one. The parameters of the initial relative orbit are $\eta=0.5$ and $D=1.0$.
We assume that the virial mass of the DM halo, $M_{\rm vir}$, doubles through a major merger and the virial radius, $r_{\rm vir}$, grows with the relation, $r_{\rm vir} \propto M_{\rm vir}^{1/3}$. 
Following the setup of the simulation of the first merger (labeled `nr'), 80\% of the total mass is expected to be contained in the virial radius of the product after the merger. 
Equation (\ref{eq:epsilon}) determines the softening length, $\epsilon$, with the control parameter, $A=4$ (for models with $\alpha=1.0$) and 1 (for models with $\alpha=1.5$). 
In the simulations of the third (labeled `3rd') and fourth (labeled `4th') mergers, the remnant of the previous merger simulation and the copied identical one are collided in the same way. 

\input{./fig10.tex}
The upper panel of Figure \ref{fig:multiple_mergers_a1.5} illustrates the density profiles of the products of consecutive mergers in which the original progenitors are characterized by the model with the central density slope of $\alpha=1.5$ (see Equation \ref{eq:inside}). These consecutive merger simulations show that the shape of the density profile changes through mergers because of the radial dependence in the efficiency of the density enhancement (see Figure \ref{fig:dns_enh}). 
The lower panel depicts that the logarithmic density slope becomes higher in the centre, i.e. shallowing the central cusp, in each merging event. 
After the fourth merger, the slope reaches approximately $-1.2$ at the centre. 
Assuming that the virial mass of the progenitors in the first merger is $7.14 \times 10^{-7} M_{\rm \odot}$, the inner slope is expected to be -1.47 based on the empirical law for the logarithmic central density slope provided by Equation 2 in I14.
The predicted median value of the inner slope of primordial DM haloes with 16 times larger virial mass, -1.33, is steeper than the result of the controlled simulations, -1.2, but it is in the range of 25-75\% (see Figure 7 of I14). 

\input{./fig11.tex}
Figure \ref{fig:multiple_mergers_a1.0} shows the density profiles (upper panel) and their logarithmic density slope profiles (lower panel) of the products of consecutive mergers where the original progenitors were given by the NFW profiles with the central density slope of $\alpha=1.0$. We find that the central cusp becomes shallower in each merger, similar to the case of the model with $\alpha=1.5$. While the density enhancement is larger for the NFW profile, its central density slope is less susceptible to the impact of major mergers because the density enhancement is more uniform (i.e., less radial dependent) in the central regions ($r < 0.2 r_{\rm vir}$). 
Our results suggest that, although the density enhancement is still significant and hence not in quasi-final state, the slope of the NFW profile is more resilient to the impact of major mergers.

These simulation results support the idea that the NFW profile could be an attractor solution of the density structure of DM haloes \citep[e.g.][]{2004ApJ...614...17G}. 
Cosmological simulations with the cutoff in the small-scale matter power spectrum predict that DM haloes near the free streaming scale first emerge with steeper central cusps than the NFW profile. The dynamical impacts of repeated major mergers make the steep central density cusps shallower (as shown in Figure~\ref{fig:multiple_mergers_a1.5}) and drive the central density slope to approach the {\it universal} form characterized by the NFW profile.
Since the lower mass limit of DM haloes is expected to be around $10^{-6} M_{\rm \odot}$ for the DM particle mass of 100~GeV, DM haloes should have experienced the growth by $10^4$ times in their mass, which corresponds to approximately 13-14 major mergers. 
Once reaching the NFW profile, the central density slope becomes highly resilient to the impacts of mergers (as shown in Figure~\ref{fig:multiple_mergers_a1.0}). 

\section{Summary and discussion}
\label{sec:sum}

Upcoming observations may detect the extremely ancient structures, such as the first stars, black holes, and galaxies, forming in primordial DM haloes in the early universe. 
Primordial DM haloes are also possible sites of DM annihilations because of their high density and abundance. 
However, the structure of the primordial DM haloes are sensitive to the free streaming scales of DM particles. 
In this work, we analyse the data from the cosmological simulations of primordial DM haloes performed in I14 in order to characterize the role of merger parameters on the structure and evolution of the primordial DM haloes. We will then use a suite of controlled, high-resolution $N$-body simulations to understand dynamical processes that govern the evolution of the density structure in the primordial DM haloes. Our main findings are summarized below.

\begin{enumerate}

\vspace{2mm}
\item 
We test the hypothesis of \citet[][I14]{2014ApJ...788...27I} that the central cusps contained in primordial DM haloes become shallower through mergers by analysing the data from the cosmological simulation with the cutoff in the power spectrum performed in I14 and find that the density cusps contained in the centre of primordial DM haloes which experienced mergers are shallower than those of DM haloes which have not experienced mergers.  
We also find that primordial DM haloes near the free streaming scale have more radial merger orbits and the fraction of major mergers is higher compared with DM haloes in cosmological simulations without the cutoff. 

\vspace{2mm}
\item By analysing a suite of idealized collisionless binary merger simulations, we find that the dynamical impacts depend on the slope of the cusps contained in merger progenitors. For example, the central density does not double through a major merger between DM haloes with very steep cusps ($\alpha \geq 1.5$) like ones contained in primordial DM haloes near the free streaming scale, while major mergers between DM haloes which initially follow the NFW profiles ($\alpha=1.0$), like galaxy-sized DM haloes, double the central density.

\vspace{2mm}
\item 
We show that mergers between two DM haloes lead to the process of violent relaxation and to the strong accelerations in the direction of the collision between the two haloes. A part of DM particles near their centre escapes from their potential well and creates a significant symmetrical density enhancement in all simulation runs. Moreover, in DM haloes with very steep density cusps ($\alpha \geq 1.5$), some fraction of DM particles also escapes from the centre isotropically, because these systems are dynamically hotter than ones with shallower cusps in the central region.

\vspace{2mm}
\item 
Results of our simulations provide new insights into the role of dynamical processes that shape the central cusps of primordial DM haloes. While the density in the outer regions almost doubles in all equal mass mergers, the enhancement in the central density is smaller in a major merger between DM haloes with steeper cusps compared to that of the shallower NFW profile. These results suggest the slope of the density cusp in the primordial DM haloes becomes shallower through repeated major mergers in the early universe.

\vspace{2mm}
\item 
We developed a simple model to describe the evolution of DM haloes through a series of consecutive major mergers and tested the validity of such a model using a series of controlled simulations of multiple major mergers. Both of our simple estimation and controlled simulations show that consecutive major mergers between DM haloes near the free streaming scale reproduce the density structure of DM haloes with larger virial masses. The slope of the central density cusps of DM haloes becomes shallower in each merger, indicating that the central density structure depends on the dynamical processes of major mergers, such as violent relaxation and phase-space mixing.

\vspace{2mm}
\item 
We find that the central density slope of the NFW profile is more resilient to the dynamical impacts of consecutive major mergers compared to ones with steeper cusps because the density enhancement is more uniform within the virialized region of DM halos. This lends support to the idea that the NFW profile could be an attractor solution. Our work shed new insights into the impact of major mergers on the structure of DM halos as well as the prevalence of the {\it universal} form of DM density structure on all cosmological scales.

\vspace{2mm}
\item
We emphasize that the structural parameters of merger progenitors adopted in our controlled simulations are quite different from those in previous studies and lead to different conclusions. 
Most of previous such studies have focused on the impact of mergers on the galactic-size DM haloes with relatively higher concentration ($c \sim 10$) and the NFW profile which corresponds to the model of $\alpha=1.0$ (see Equations \ref{eq:nfw} and \ref{eq:inside}). These works have concluded that the slope of the steepest cusps is well-preserved through gas-less mergers, and mergers do not significantly modify the inner structure of DM haloes. Our present work, on the other hand, focuses on primordial DM haloes with a considerably smaller concentration parameter, $c = 2$, and finds that major mergers plays a significant role in determining their central density structure. 

\end{enumerate}

Our work highlights the importance of studying the dynamical impacts of mergers on DM haloes with the low concentration parameter, which has not been studied in the literature. Future work should focus on the role of finite DM streaming scale on the formation and evolution of first stars, black holes, and galaxies using cosmological hydrodynamical simulations.  Such work should shed light on the structure formation during the Dark Ages and provide necessary theoretical support for interpreting upcoming observations with {\it JWST}, {\it GMT}, {\it TMT}, and {\it E-ELT}.

\section*{Acknowledgments}
We thank Max-Planck-Institut f{\"u}r Astrophysik for hospitality where this work was conceived, and Andreas Burkert, Masao Mori, Simon White, and an anonymous referee for useful comments.
Numerical computations were carried out on Aterui supercomputer at Center for Computational Astrophysics, CfCA, of National Astronomical Observatory of Japan, the K computer at the RIKEN Advanced Institute for Computational Science (Proposal numbers hp150226 and hp150263) and HA-PACS at the Center for Computational Sciences at University of Tsukuba. 
GO has been partially supported by the DFG cluster of excellence `Origin and Structure of the Universe' (www.universe-cluster.de). 
DN was supported in part by NSF grant AST-1412768 and the Research Corporation. TI has been funded by MEXT HPCI STRATEGIC PROGRAM and MEXT/JSPS KAKENHI Grant Number 15H01030. 

\bibliographystyle{mn2e}
\bibliography{./ref}

\label{lastpage}
\end{document}

%% file: fig1.tex
\begin{figure}
  \centering 
   \includegraphics[width=90mm]{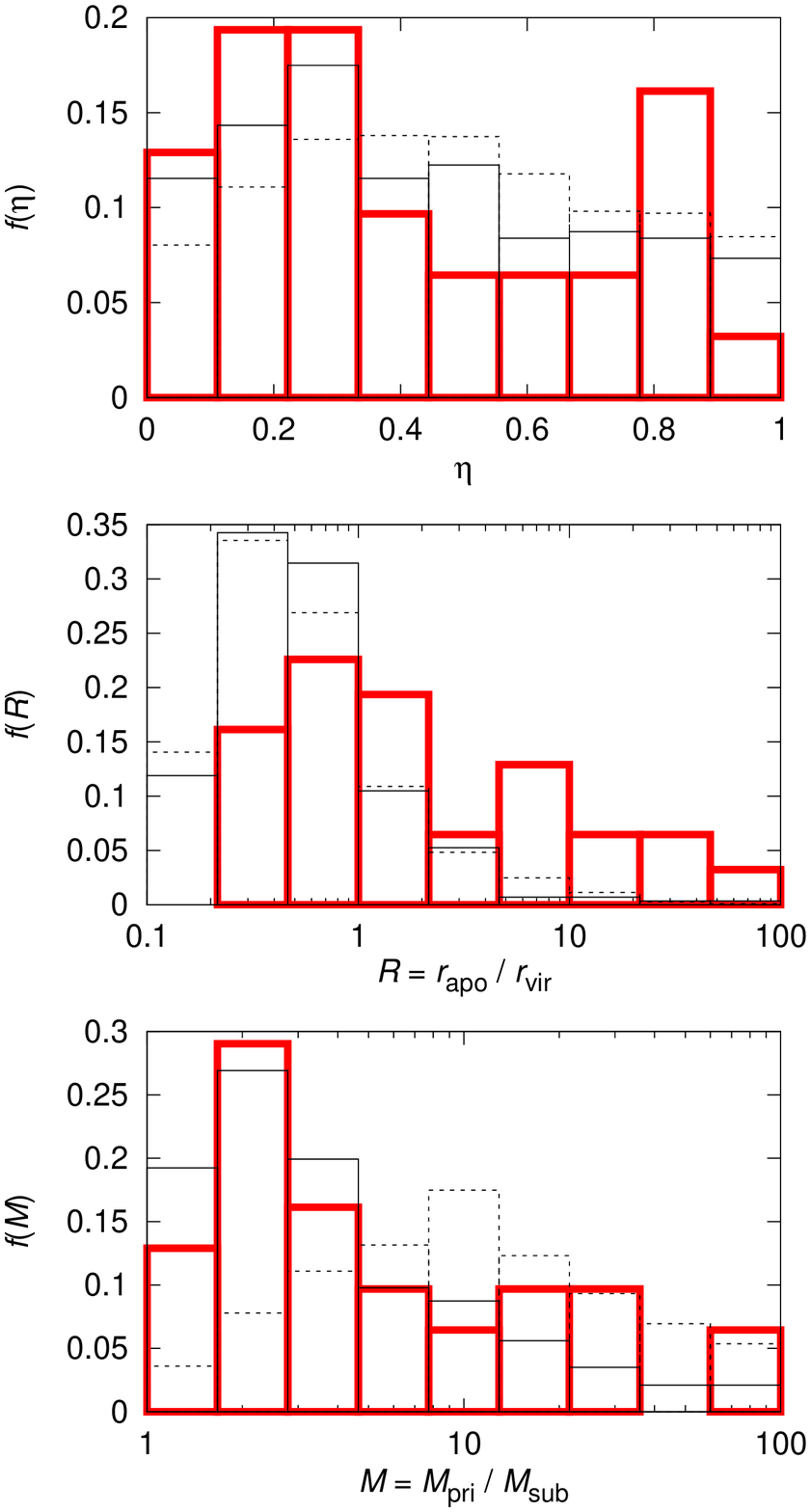}
     \caption{
       Distributions of the orbital circularity, $\eta$ (top), apocentre scaled by the virial radius of the primal haloes, $R \equiv r_{\rm apo} / r_{\rm vir}$ (middle) and mass ratio between the primal- and subhaloes, $M \equiv M_{\rm vir, pri} / M_{\rm vir, sub}$ (bottom).  
         Mergers in which the virial mass of the primal halo with $M_{\rm vir, pri} \geq 7.14 \times 10^{-7} M_{\rm \odot}$ are included in the analysis. 
         Red and black lines represent the results for the model A and B, respectively. 
         The solid and dotted lines correspond to the minimum subhalo masses of $M_{\rm sub, min} = 7.14 \times 10^{-7}$ and $7.14 \times 10^{-8} M_{\rm \odot}$, respectively.}
       \label{fig:mrg_params}
\end{figure}

%% file: fig2.tex
\begin{figure}
  \centering 
   \includegraphics[width=85mm]{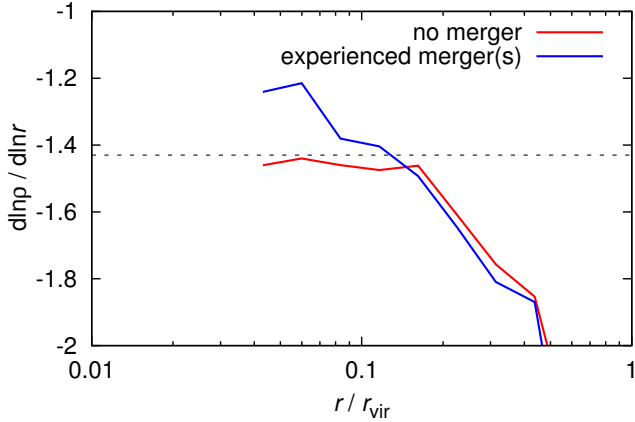}
     \caption{
       Radial profile of the logarithmic density slope of primordial DM haloes 
       in the model A, the case with the cutoff.   
       The radial bins are scaled by the virial radius of individual haloes. 
       Density structures are derived by stacking DM haloes which have virial masses $1 \times 10^{-6} \leq M_{\rm vir}/M_{\rm\odot} \leq 4 \times 10^{-6}$ at $z=32$.
       Red and blue lines represent primordial DM haloes which have not and have experienced mergers by $z=32$. 
       Black line corresponds to the prediction given by the fitting formula in Equation (2) of I14. 
       \label{fig:slope_cosmo}
     }
\end{figure}

%% file: fig3.tex
\begin{figure*}
  \centering 
   \includegraphics[width=150mm]{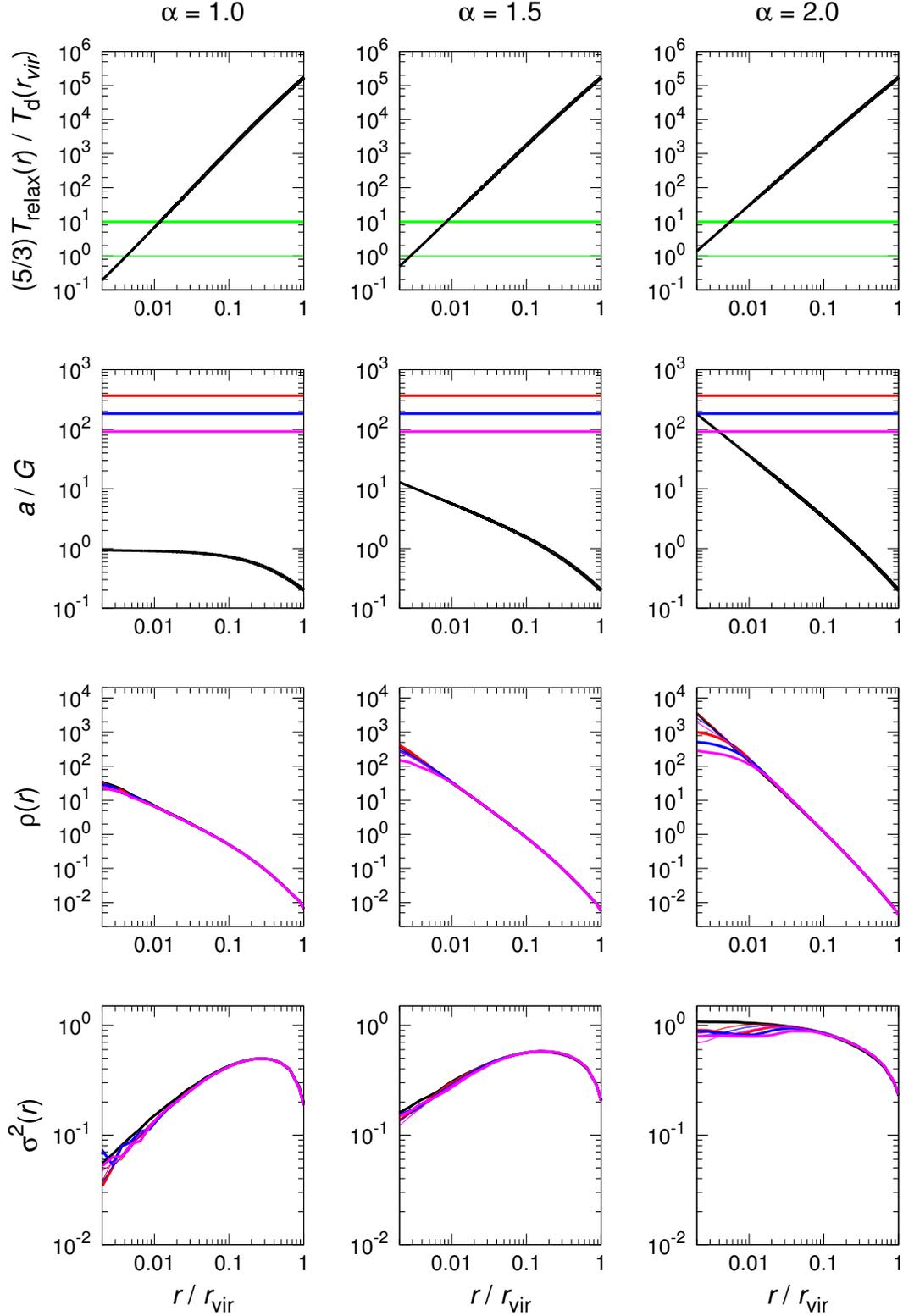}
     \caption{
       Stability tests of isolated $N$-body systems and required force resolutions. 
       The radial bins are scaled by the virial radius of each system. 
       Left, center and right panels show the results for the models of $\alpha = 1.0 \ ({\rm NFW}), 1.5$ and $2.0$. 
       Red, blue and magenta lines correspond to the cases of $A = 1, 2$ and $4$, respectively. 
       16,777,216 particles are employed in each run. 
       Descriptions of panels in respective rows: 
       (1st row) Black lines show the two-body relaxation time normalized by the free-fall time measured at $r_{\rm vir}$, $T_{\rm d}(r_{\rm vir})$. Thin and thick green lines correspond to $T_{\rm int} = 1$ and $10 T_{\rm d}(r_{\rm vir})$, respectively. 
       (2nd row) Black lines represent the expected gravitational acceleration normalized by $G$, $M(<r)/r^2$. Colored lines show the characteristic acceleration defined by Equation (\ref{eq:achar}). 
       (3rd row) Density profiles of isolated $N$-body systems. Black lines are the initial conditions. Colored thin and thick lines represent snapshots at $t = 1$ and $10 T_{\rm d}(r_{\rm vir})$. 
       (4th row) Profiles of velocity dispersion. Each line expresses the same snapshot as shown in the third row. 
       \label{fig:res_study}
     }
\end{figure*}

%% file: fig4.tex
\begin{figure}
  \centering 
   \includegraphics[width=85mm]{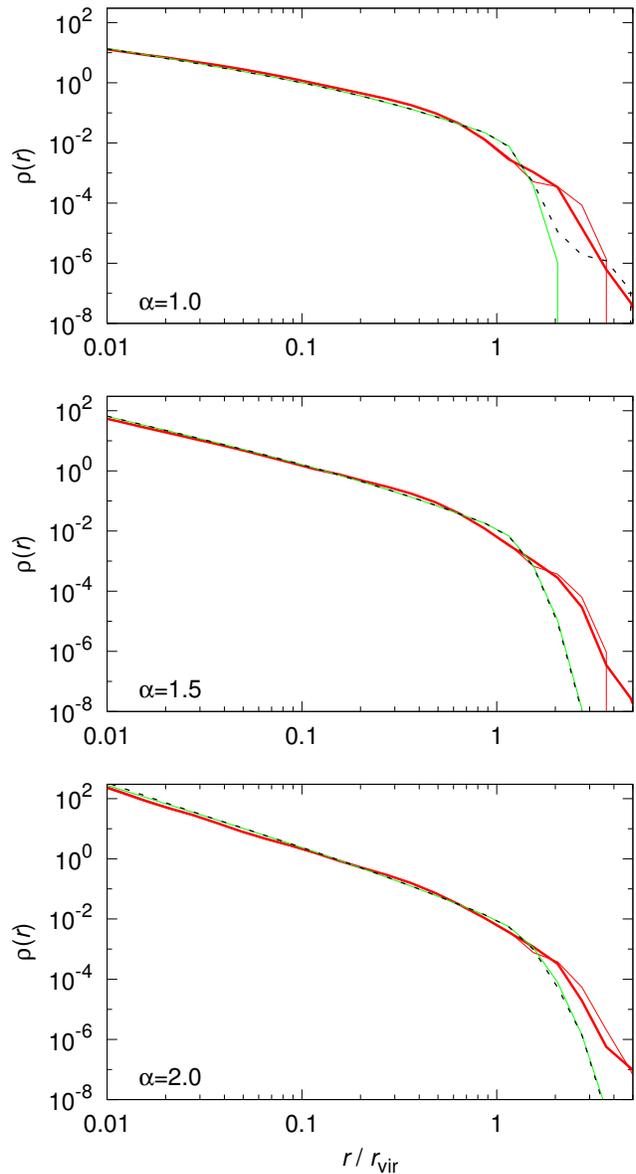}
     \caption{
       Spherically averaged density profile of the merged haloes, $\rho(r)$, in runs with $\eta = 0.5$ and D=1.0. 
         The radial bins are scaled by the virial radius of the merger progenitors, $r_{\rm vir}$. 
       Top, middle and bottom panels show the results for the models of $\alpha = 1.0, 1.5$ and $2.0$. 
       Thick and thin red lines represent the merger remnant at $t = 5$ and $10 T_{\rm d}(r_{\rm vir})$, respectively. 
       The merged haloes have reached the quasi-equilibrium state. 
       Green lines express the initial state of the merger progenitor. 
       Black dashed lines are the snapshot at $t = 10 T_{\rm d}(r_{\rm vir})$ in iso-runs and test the stability of the progenitors. 
       The density values of green and black dashed lines are doubled. 
       \label{fig:dns_prf}
     }
\end{figure}

%% file: fig5.tex
\begin{figure*}
   \includegraphics[width=180mm]{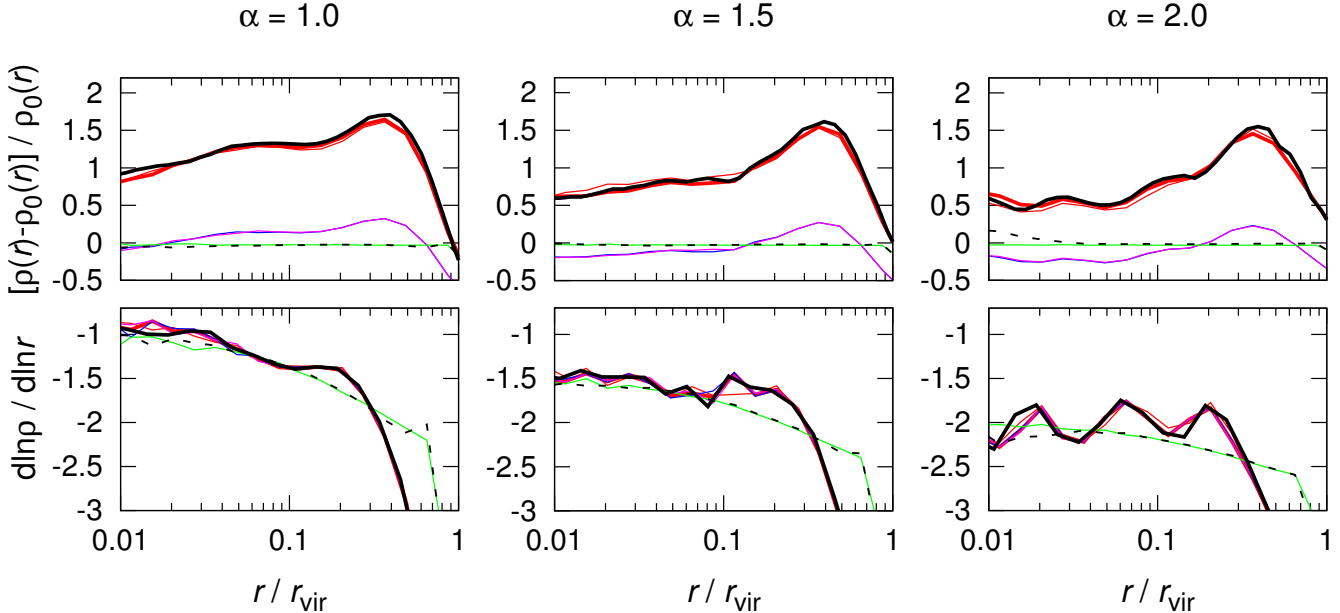}
     \caption{
       Enhancement in the radial density profile, $[\rho(r) - \rho_0(r)] / \rho_0(r)$ 
       (upper panels) and the logarithmic slope of the density profile, ${\rm d} \ln{\rho(r)} / {\rm d}\ln{r}$ (lower panels), in runs with $\eta=0.5$ and D=1.0, where $\rho_0(r)$ is the analytical expression of the radial density profile of the progenitor. 
       The radial bins are scaled by the virial radius of the merger progenitors, $r_{\rm vir}$. 
       Left, middle and right panels show the results for the models of $\alpha = 1.0, 1.5$ and $2.0$. 
       Thin and thick red lines represent the merger remnants in nr-runs at $t = 5$ and $10 T_{\rm d}(r_{\rm vir})$, respectively. 
       Merger remnants have reached the quasi-equilibrium states. 
       Black solid lines depict the results of hr-runs at $t = 10 T_{\rm d}(r_{\rm vir})$. 
       Results are numerically converged. 
       Blue and magenta lines show the contributions from each progenitor. 
       Green lines are the initial state of the progenitor. 
       Black dashed lines are the snapshot at $t = 10 T_{\rm d}$ in iso-runs and express the stability of the progenitors. 
       \label{fig:dns_enh}
     }
\end{figure*}

%% file: table1.tex
\begin{table}
\begin{center}
\caption{
Summary of controlled simulation runs to test the stability of $N$-body systems (iso-runs) and the dependence on the internal structure of merger progenitors (nr- or hr-runs). 
Description of columns: 
(1) Name of simulation runs. 
Runs labeled `iso' simulate the isolated systems to test their stability. 
Runs labeled `nr' or `hr' are merger simulations with normal or higher resolutions. 
(2) Initial logarithmic slope of the central density cusp, $\alpha$. 
(3) Orbital circularity, $\eta$, defined as the ratio between the angular momentum, $L$, and that of the circular orbit, $L_{\rm c}$, i.e. $\eta \equiv L / L_{\rm c}$. 
(4) Parameter to determine the initial separation between the centres of two systems, $D$. The initial separation, $d$, is given by $d = D r_{\rm vir}$. 
(5) Number of particles in each system, $N$. Total number of particles, $N_{\rm tot}$, is $2 N$ in nr- or hr-runs. In iso-runs, $N_{\rm tot} = N$. 
}
\begin{tabular}{ccccc}
Run & $\alpha$ & $\eta$ & $D$ & $N$ \\ 
(1) & (2) & (3) & (4) & (5) \\
\hline 
a1.0-iso & 1.0 & - & - & 16,777,216 \\
a1.0-nr & 1.0 & 0.5 & 1.0 & 16,777,216  \\
a1.0-hr & 1.0 & 0.5 & 1.0 & 104,857,600  \\
a1.5-iso & 1.5 & - & - & 16,777,216 \\
a1.5-nr & 1.5 & 0.5 & 1.0 & 16,777,216  \\
a1.5-hr & 1.5  & 0.5 & 1.0 & 104,857,600 \\
a2.0-iso & 2.0 & - & - & 16,777,216 \\
a2.0-nr & 2.0 & 0.5 & 1.0 & 16,777,216  \\
a2.0-hr & 2.0  & 0.5 & 1.0 & 104,857,600 \\
\hline 
\end{tabular}
\label{tab:run_str}
\end{center}
\end{table}

%% file: fig6.tex
\begin{figure}
  \centering 
   \includegraphics[width=85mm]{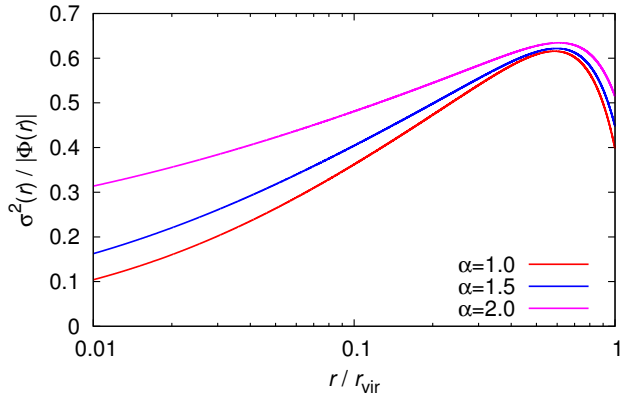}
     \caption{
       Radial profile of the pseudo virial ratio defined by $\sigma^2(r)/|\Phi(r)|$, where $\sigma(r)$ and $\Phi(r)$ mean the radial profiles of three-dimensional velocity dispersion and gravitational potential. 
       The radial bins are scaled by the virial radius of the systems, $r_{\rm vir}$. 
       Red, blue and magenta lines represent the initial state of the progenitors with $\alpha = 1.0, 1.5$ and $2.0$, respectively.
       \label{fig:pseudo_vir_ratio}
     }
\end{figure}

%% file: fig7.tex
\begin{figure}
  \centering 
   \includegraphics[width=80mm]{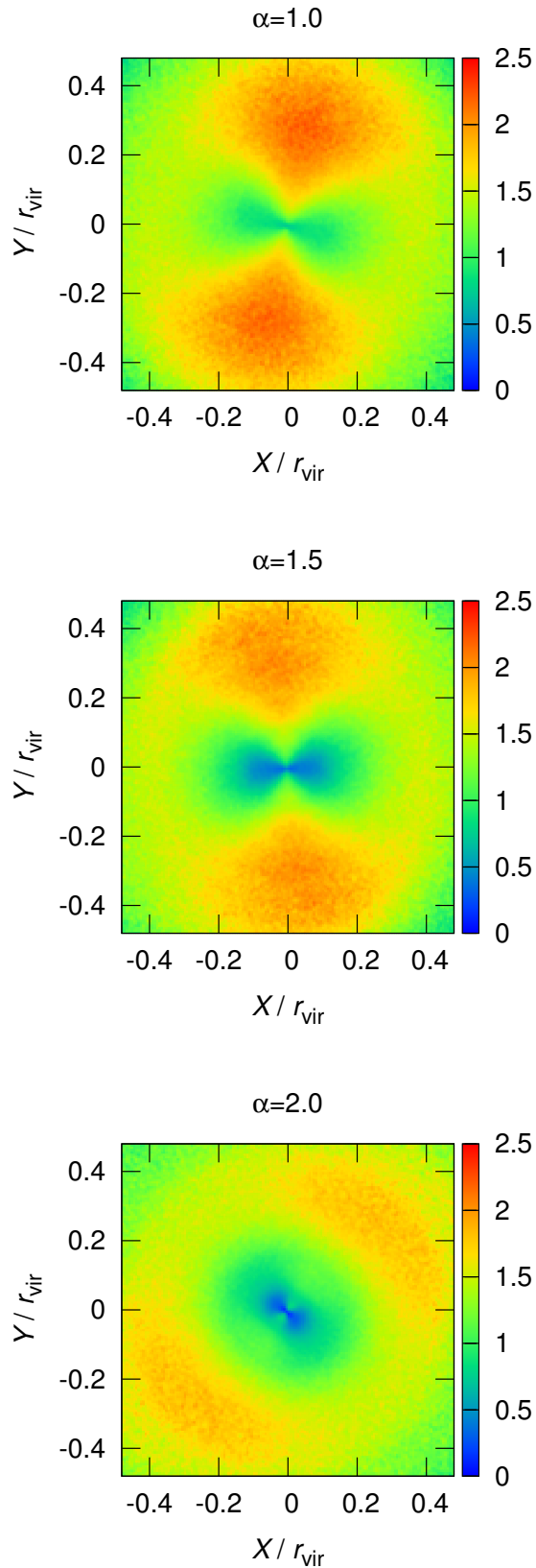}
   \caption{
     Distribution of the enhancement in the column density of the merger remnants in hr-runs. 
     The snapshots at $10 T_{\rm d}(r_{\rm vir})$ are shown. 
     The color bar represents the enhancement in the column density, $[\Sigma - \Sigma_0]/\Sigma_0$, where $\Sigma_0$ is the initial column density of the progenitor at given position. 
     Spatial coordinates are scaled by the virial radius of the merger progenitor, $r_{\rm vir}$. 
       \label{fig:dns_field}
     }
\end{figure}

%% file: table2.tex
\begin{table}
\begin{center}
\caption{
Summary of controlled simulation runs to study the dependence on the orbital parameters of mergers. The meaning of symbols is described in Table \ref{tab:run_str}. 
}
\begin{tabular}{ccccc}
Run & $\alpha$ & $\eta$ & $D$ & $N$ \\ 
(1) & (2) & (3) & (4) & (5) \\
\hline 
a1.5-h0.1 & 1.5 & 0.1 & 1.0 & 16,777,216  \\
a1.5-h0.3 & 1.5 & 0.3 & 1.0 & 16,777,216  \\
a1.5-h0.7 & 1.5 & 0.7 & 1.0 & 16,777,216  \\
a1.5-h0.9 & 1.5 & 0.9 & 1.0 & 16,777,216  \\
a1.5-d0.5 & 1.5 & 0.5 & 0.5 & 16,777,216  \\
a1.5-d2.0 & 1.5 & 0.5 & 2.0 & 16,777,216  \\
\hline 
\end{tabular}
\label{tab:run_orb}
\end{center}
\end{table}

%% file: fig8.tex
\begin{figure}
  \centering 
   \includegraphics[width=85mm]{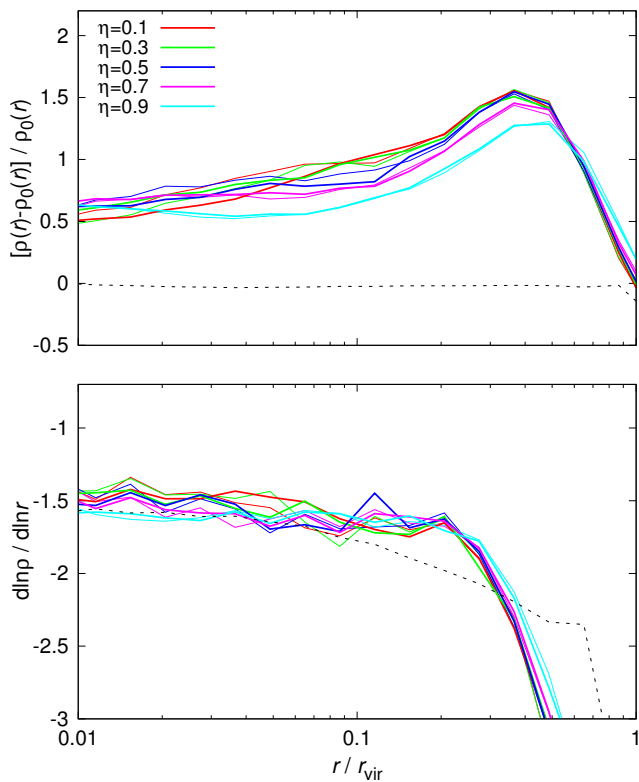}
     \caption{
       Same as Figure \ref{fig:dns_enh}, but examining the dependence on orbital circularity, $\eta$. 
       Red, green, blue, magenta and cyan lines are the merger remnants in runs named a1.5-h0.1, a1.5-h0.3, a1.5-nr, a1.5-h0.7 and a1.5-h0.9, respectively. 
       Thin and thick lines represent snapshots at $t = 5$ and $10 T_{\rm d}(r_{\rm vir})$. 
       Black dashed line is the snapshot at $t = 10 T_{\rm d}(r_{\rm vir})$ in the a1.5-iso run.  
       \label{fig:dns_enh_eta}
     }
\end{figure}

%% file: fig9.tex
\begin{figure}
  \centering 
   \includegraphics[width=85mm]{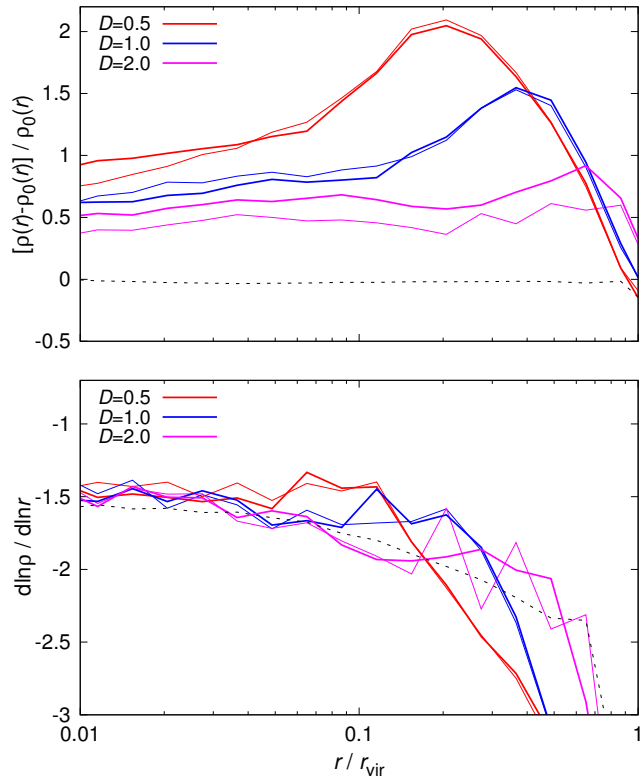}
     \caption{
       Same as Figure \ref{fig:dns_enh}, but examining the dependence on the initial separation between two merger progenitors, $d = D r_{\rm vir}$, where the dimensionless parameter, $D$, controls the separation size and $r_{\rm vir}$ is the virial radius of the progenitors. 
       Red, blue and magenta lines are the merger remnants in runs named a1.5-d0.5, a1.5-nr and a1.5-d2.0, respectively. 
       Thin and thick lines represent snapshots at $t = 5$ and $10 T_{\rm d}(r_{\rm vir})$. 
       Black dashed line is the snapshot at $t = 10 T_{\rm d}(r_{\rm vir})$ in the a1.5-iso run. 
       \label{fig:dns_enh_r}
     }
\end{figure}

%% file: table3.tex
\begin{table}
\begin{center}
\caption{Model application to the results of I14. 
Description of columns: 
(1) Radius scaled by the virial radius of DM haloes with $2 \times 10^{-6} M_{\rm \odot}$, $r_{\rm vir, 6}$. 
(2) Density of DM haloes with $2 \times 10^{-6} M_{\rm \odot}$ measured at $r$ (from I14). 
(3) Radius scaled by the virial radius of DM haloes with $5 \times 10^{-5} M_{\rm \odot}$, $r_{\rm vir, 5}$.
(4) Density of DM haloes with $5 \times 10^{-5} M_{\rm \odot}$ measured at $r$ (from I14). 
(5) Model prediction for the density of DM haloes with $5 \times 10^{-5} M_{\rm \odot}$ measured at $r$. 
The density is given in the unit of $M_{\rm \odot} {\rm pc}^{-3}$.
}
\begin{tabular}{ccccc}
$r/r_{\rm vir, 6}$ & $\rho_{\rm I14, 6}(r)$ & $r/r_{\rm vir, 5}$ & $\rho_{\rm I14, 5}(r)$ & $\rho_{\rm model, 5}(r)$ \\ 
(1) & (2) & (3) & (4) & (5) \\
\hline 
0.06 & $\sim 10$ & 0.02 & $\sim 60$ & 50 - 75 \\
0.9  & $\sim 0.1$ & 0.3  & $\sim 1$    & 1.6 - 3.2 \\
\hline 
\end{tabular}
\label{tab:app}
\end{center}
\end{table}

%% file: table4.tex
\begin{table}
\begin{center}
\caption{
Summary of controlled simulation runs to study the dynamical impacts of consecutive mergers. The meaning of symbols is described in Table \ref{tab:run_str}. 
}
\begin{tabular}{ccccc}
Run & $\alpha$ & $\eta$ & $D$ & $N$ \\ 
(1) & (2) & (3) & (4) & (5) \\
\hline 
a1.0-2nd & 1.0 & 0.5 & 1.0 &  33,554,432 \\
a1.0-3rd & 1.0 & 0.5 & 1.0 &  67,108,864 \\
a1.0-4th & 1.0 & 0.5 & 1.0 &  134,217,728\\
a1.5-2nd & 1.5 & 0.5 & 1.0 &  33,554,432 \\
a1.5-3rd & 1.5 & 0.5 & 1.0 &  67,108,864 \\
a1.5-4th & 1.5 & 0.5 & 1.0 &  134,217,728\\
\hline 
\end{tabular}
\label{tab:run_cum}
\end{center}
\end{table}

%% file: fig10.tex
\begin{figure}
  \centering 
   \includegraphics[width=85mm]{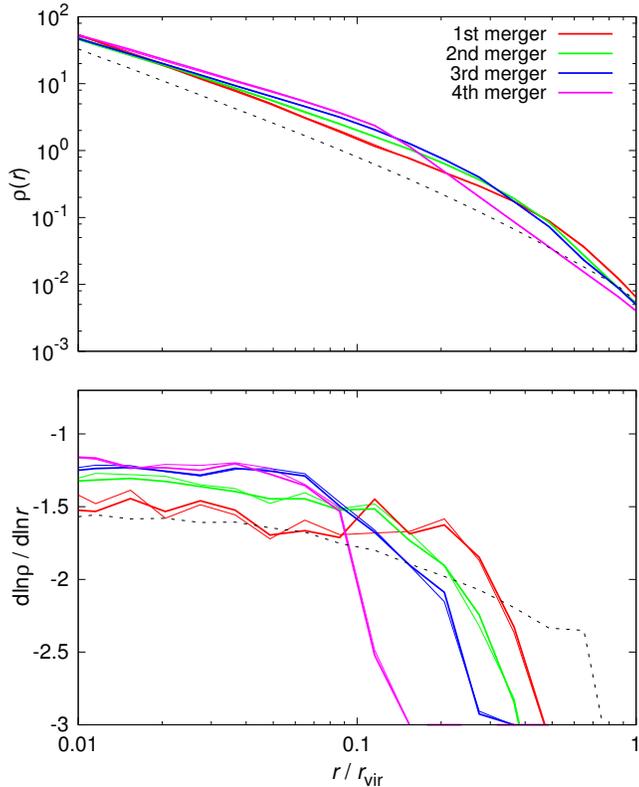}
     \caption{
         Density profiles (upper panel) and logarithmic density slope (lower panel) of the merger remnants of progenitors with the central density slope of $\alpha=1.5$.
         Red, green, blue and magenta lines are the remnants of the first (a1.5-nr), second (a1.5-2nd), third (a1.5-3rd) and fourth (a1.5-4th) major mergers, respectively. 
         Thin and thick lines represent snapshots at $t = 5$ and $10 T_{\rm d}(r_{\rm vir})$. 
         Black dashed line is the snapshot at $t = 10 T_{\rm d}(r_{\rm vir})$ in the a1.5-iso run. 
         Radial bins are scaled by the virial radius of the merger progenitors, $r_{\rm vir}$, in each run. 
       \label{fig:multiple_mergers_a1.5}
     }
\end{figure}

%% file: fig11.tex
\begin{figure}
  \centering 
   \includegraphics[width=85mm]{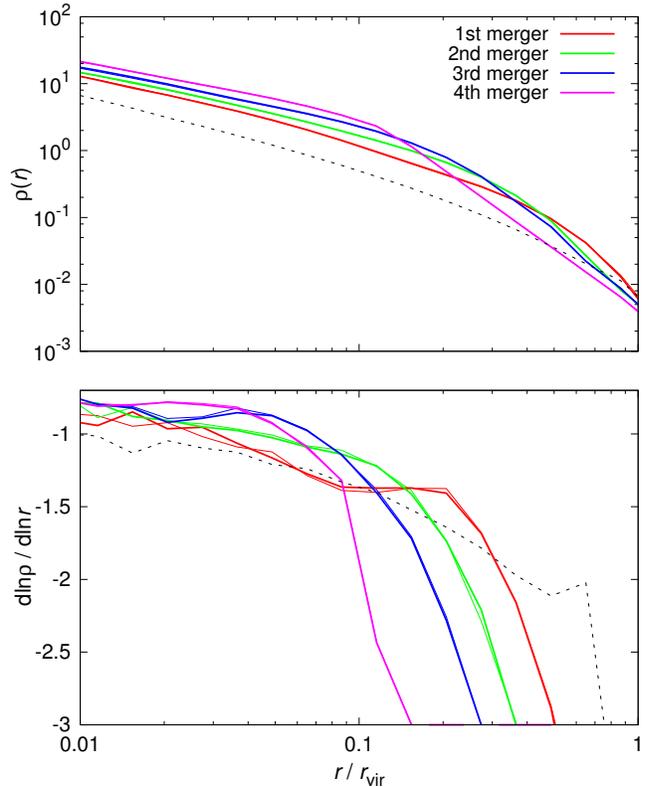}
     \caption{
         Same as Figure \ref{fig:multiple_mergers_a1.5}, but for the central density slope $\alpha=1.0$ for the progenitors. 
         Red, green, blue and magenta lines are the remnants of the first (a1.0-nr), second (a1.0-2nd), third (a1.0-3rd) and fourth (a1.0-4th) major mergers, respectively. 
       \label{fig:multiple_mergers_a1.0}
     }
\end{figure}